\documentclass[12pt,preprint]{aastex}
\usepackage{emulateapj5}
\usepackage{apjfonts}
\usepackage{natbib}

\input psfig.tex

\def\aa{{A\&A}}

\def\aj{{AJ}}
\def\al{$\alpha$}
\def\bet{$\beta$}

\def\annrev{{ARA\&A}}
\def\apj{{ApJ}}
\def\apjs{{ApJS}}
\def\asec{$^{\prime\prime}$}

\def\cc{cm$^{-3}$}

\def\cc{cm$^{-3}$}

\def\etal{{et al. }}

\def\kms{km s$^{-1}$}
\def\lamb{$\lambda$}
\def\lax{{$\mathrel{\hbox{\rlap{\hbox{\lower4pt\hbox{$\sim$}}}\hbox{$<$}}}$}}
\def\gax{{$\mathrel{\hbox{\rlap{\hbox{\lower4pt\hbox{$\sim$}}}\hbox{$>$}}}$}}
\def\simlt{\lower.5ex\hbox{$\; \buildrel < \over \sim \;$}}
\def\simgt{\lower.5ex\hbox{$\; \buildrel > \over \sim \;$}}
\def\lum{erg s$^{-1}$}
\def\mbh{{$M_{\rm BH}$}}

\def\mnras{{MNRAS}}

\def\pasp{{PASP}}

\def\percm2{cm$^{-2}$}

\def\solmass{$M_\odot$}
\def\oii{[\ion{O}{2}]}

\def\hi{\ion{H}{1}}
\def\hii{\ion{H}{2}}
\def\oiii{[\ion{O}{3}]}

\def\nii{[\ion{N}{2}]}

\def\sii{[\ion{S}{2}]}

\def\lhal{$L_{{\rm H}\alpha}$}
\def\lbol{$L_{{\rm bol}}$}
\def\ledd{$L_{{\rm Edd}}$}
\def\sigg{$\sigma_g$}
\def\sigs{$\sigma_*$}

\slugcomment{To appear in {\it The Astrophysical Journal}}
\shorttitle{IONIZED GAS IN BULGES}
\shortauthors{HO}

\begin{document}

\title{Origin and Dynamical Support of Ionized Gas in Galaxy Bulges}

\author{Luis C. Ho}

\affil{The Observatories of the Carnegie Institution of Washington, 813 Santa 
Barbara Street, Pasadena, CA 91101}

\begin{abstract}
We combine ionized gas (\nii\ \lamb6583) and stellar central velocity 
dispersions for a sample of 345 galaxies, with and without active galactic 
nuclei (AGNs), to study the dynamical state of the nuclear gas and its 
physical origin.  The gas dispersions strongly correlate with the stellar 
dispersions over the velocity range of $\sigma \approx 30-350$ \kms, such that 
$\sigma_g/\sigma_* \approx 0.6-1.4$, with an average value of 0.80.  These 
results are independent of Hubble type (for galaxies from E to Sbc), presence 
or absence of a bar, or local galaxy environment.  For galaxies of type 
Sc and later and that have \sigs\ \lax\ 40 \kms, the gas seems to have a 
minimum threshold of \sigg\ $\approx$ 30 \kms, such that \sigg/\sigs\ always 
exceeds 1.  Within the sample of AGNs, \sigg/\sigs\ increases with nuclear 
luminosity or Eddington ratio, a possible manifestation of AGN feedback 
associated with accretion disk winds or outflows.  This extra source of 
nongravitational line broadening should be removed when trying to use \sigg\ 
to estimate \sigs.  We show that the mass budget of the narrow-line region 
can be accounted for by mass loss from evolved stars.  The kinematics of the 
gas, dominated by random motions, largely reflect the velocity field of 
the hot gas in the bulge.  Lastly, we offer a simple explanation for the 
correlation between line width and line luminosity observed in the narrow-line 
region of AGNs.
\end{abstract}


\section{Introduction}

Warm ($\sim 10^4$ K), ionized gas is pervasive in the central regions of 
nearby galaxies of all morphological types.  Easily detected through optical 
emission lines, either by imaging or spectroscopy, this component of the 
interstellar medium has served as a diagnostic of the physical conditions in 
galactic nuclei, including their source of excitation and chemical 
abundances.  The optical line emission also provides a powerful tracer of the 
nuclear kinematics of the galaxy, in efforts to probe the mass distribution of 
the bulge, to detect central dark massive objects, and to study substructure 
such as kinematically decoupled cores.

What is the dynamical state of the nuclear gas?  How does it relate to the 
dynamics of the stars?  Is the gas primarily rotationally supported, or do 
random motions also contribute?  While these issues have been addressed in the 
past, they have been based either on investigations of individual objects or 
small, restricted samples.  A number of authors have noted, for example, that 
in the bulges of disk (lenticular and spiral) galaxies, the gas often rotates 
slower than the stars (Fillmore et al. 1986; Kent 1988; Kormendy \& Westpfahl 
1989; Bertola et al. 1995; Cinzano et al. 1999; Pignatelli et al. 2001).  The 
same phenomenon has been found in some elliptical galaxies (e.g., Caldwell 
1984; Caldwell et al. 1986).  The physical origin for this effect, however, is 
not entirely clear.  An interesting possibility is that the gas experiences 
nonnegligible pressure support (e.g., Kent 1988; Cinzano \& van~der~Marel 
1994; Bertola et al. 1995; Fisher 1997), an idea consistent with the 
detection of large central line widths in many bulge-dominated galaxies 
(e.g., Demoulin-Ulrich et al. 1984; Phillips et al. 1986; Bertola et al. 
1995).  Vega~Beltr\'an et al. (2001) compiled gas and stellar velocity 
measurements for a sample of $\sim$40 disk galaxies to examine trends with 
Hubble type.  They conclude that, with a few exceptions, the gas is 
dynamically cold, confined to a disk supported mainly by rotation.

What is the role of nongravitational forces on the gas dynamics, especially in 
objects harboring active galactic nuclei (AGNs)?  This issue has been 
investigated most thoroughly in the context of the narrow-line region (NLR) in 
radio (Heckman et al. 1985; Smith et al. 1990; Baum \& McCarthy 2000; 
Noel-Storr et al. 2007) and luminous Seyfert (e.g., Whittle 1992b, 1992c; 
Nelson \& Whittle 1996) galaxies.  The overall consensus from these studies is 
that the velocity field of the NLR gas largely traces the gravitational 
potential of the bulge of the host galaxy.  Super-virial motions can be found, 
but only in the minority of objects exhibiting powerful, jetlike radio 
structures, which presumably impart additional mechanical acceleration to the 
gas.  Comparing the widths of the optical nebular lines and the stellar 
velocity dispersions of a large sample of relatively luminous Seyfert~2 
galaxies, Greene \& Ho (2005a) reinforced the notion that the stellar 
potential principally governs the kinematics of the NLR, with important 
exceptions arising in cases where nuclear activity is especially strong.  To 
date, much less attention has been paid to more garden-variety AGNs such as 
low-luminosity Seyferts and LINERs, or to nuclei powered by massive stars.  

The Palomar spectroscopic survey of nearby galaxies (Ho et al. 1995, 1997a, and
references therein) offers an excellent opportunity to examine some of the 
issues mentioned above.  In addition to the existing extensive database of 
emission-line widths, a catalog of central stellar velocity dispersions is now 
available (Ho et al. 2009).  This allows a detailed comparison of the central 
velocity dispersions of the gas and stars for a large, well-defined sample of 
galaxies spanning a wide range of Hubble types and nuclear spectral classes.  
This is the goal of this paper.

\vskip 1.5cm

\section{Sample and Data}

\vskip 0.3cm
\begin{figure*}[t]
\centerline{\psfig{file=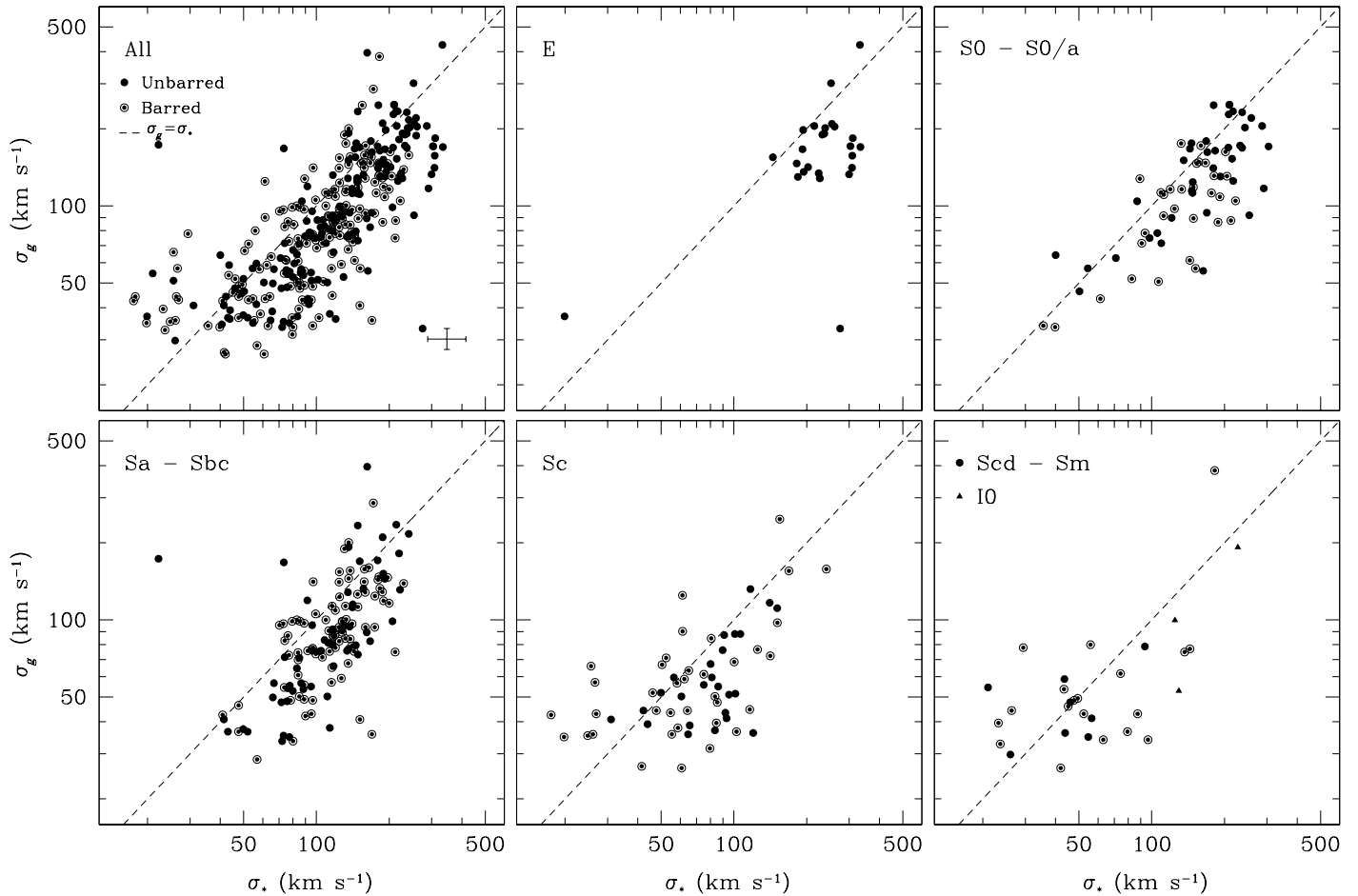,width=19.5cm,angle=0}}
\figcaption[fig1.ps]{
Distribution of gas ($\sigma_g$) versus stellar ($\sigma_*$) velocity
dispersion as a function of Hubble type.  Unbarred and barred galaxies are
plotted as filled and encircled points, respectively.  The dashed line denotes
\sigg\ = \sigs.  A typical error bar is given in the lower-right corner of the
first panel.
\label{fig1}}
\end{figure*}
\vskip 0.3cm

Our analysis draws from the nearly complete, magnitude-limited sample of 
486 bright galaxies spectroscopically surveyed using the Palomar 5-meter 
telescope.  We focus on the galaxies that have well-measured widths (FWHM) for 
\nii\ \lamb 6583, which, as explained in Ho et al. (1997a), is the line of 
choice in the Palomar survey for profile measurements.  The majority of the 
Palomar galaxies (428/486 or 88\%) now have central stellar velocity 
dispersion determined directly from the original survey data (Ho et al. 
2009).  Although many of the galaxies have preexisting velocity dispersion 
measurements in the literature, the Palomar data constitute a self-consistent, 
homogeneous set of measurements.  A significant fraction of the galaxies 
(30\%) have dispersions measured for the first time.  Of the 407 galaxies that 
have FWHM(\nii) measurements, 372 (91\%) have stellar velocity dispersions.  
If we confine our attention to the sources that  have FWHM(\nii) 
measurements with a quality rating of ``b'' or better (see Ho et al. 1997a), 
excluding a handful objects with only upper limits on their line widths, 345 
have stellar velocity dispersions.  

The above subset of 345 objects, the focus of this paper, represents fairly
the parent population of emission-line nuclei in the Palomar survey.  The 25 
E, 70 S0--S0/a, 149 Sa--Sbc, 62 Sc, and 30 Scd--Sm galaxies constitute, 
respectively, 81\%, 93\%, 96\%, 84\%, and 48\% of the parent population of 
emission-line objects.  In terms of the nuclear spectral classes, the coverage 
is excellent for the AGNs: the present subsample contains 94\% of the Seyferts, 
93\% of the LINERs, and 92\% of the transition nuclei\footnote{See Ho et al. 
(1997a) for definition of the spectral classifications.} in the parent survey.
The high completeness fractions simply reflect the high AGN detection rate in 
bulge-dominated galaxies (Ho et al. 1997b), for which velocity dispersions are 
easier to obtain.  By comparison, \hii\ nuclei are 72\% complete in the current 
sample---still a very high fraction.  

\section{The Correlation Between \sigg\ and \sigs}

\subsection{Dependence on Hubble Type}

Figure~1 shows the overall correlation between gas (\sigg) and stellar (\sigs) 
velocity dispersions, first for the entire sample, and then individually for 
several Hubble type bins.  Most of the narrow emission lines, especially near 
the core, do not deviate strongly from a simple Gaussian (see Ho et al. 1997d),
and we define \sigg\ $\equiv$ FWHM(\nii)/2.35.  Barred galaxies, taken to be 
those designated as ``SB'' or ``SAB'' in the Third Reference Catalogue of 
Bright Galaxies (de~Vaucouleurs et al. 1991), are 

\vskip 0.3cm
\begin{figure*}[t]
\centerline{\psfig{file=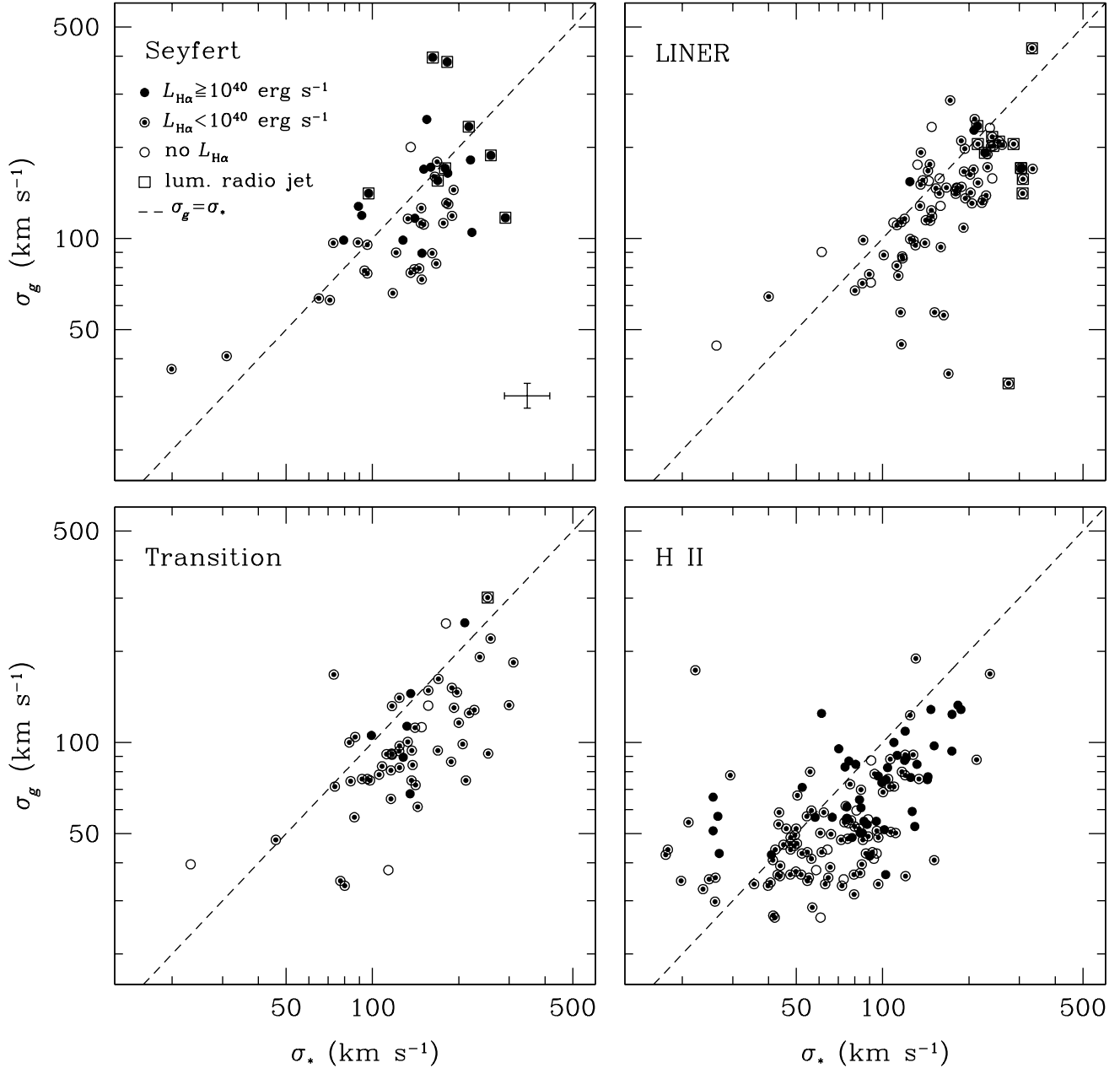,width=18.0cm,angle=0}}
\figcaption[fig2.ps]{
Distribution of gas ($\sigma_g$) versus stellar ($\sigma_*$) velocity
dispersion as a function of nuclear spectral class.  The symbols are coded
according to H\al\ luminosity.  Objects containing linear radio sources with
6~cm powers greater than $10^{21.7}$ W~Hz$^{-1}$ are marked with boxes.
The dashed line denotes \sigg\ = \sigs.  A typical error bar is given in the
lower-right corner of the first panel.
\label{fig2}}
\end{figure*}
\vskip 0.3cm

\noindent
marked for emphasis.  It is 
clear that \sigg\ correlates strongly with \sigs, albeit with considerable 
scatter.  For the sample as a whole, the Kendall's $\tau$ correlation 
coefficient (Isobe et al. 1986) is $r = 1.19$, which rejects the null 
hypothesis of no correlation 
with a probability of $P_{\rm null} < 10^{-4}$.  Both \sigg\ and \sigs\ cover 
approximately the same range\footnote{The lower limit of the velocities, 
however, is strongly influenced by the spectral resolution of the Palomar 
survey, which is $\sigma\approx 40$ \kms\ in the red and $\sim$120 \kms\ in the 
blue (Ho et al. 1997a).  Published stellar velocity dispersions often are also 
not very reliable for low-dispersion galaxies. Thus, we consider the region 
defined by \sigg\ \lax\ 40 \kms\ and \sigs\ \lax\ 40 \kms\ to be quite 
uncertain.} of velocities, from $\sim$30 to 350 \kms.  Although the 
correlation is most evident for the entire sample, the same trend replicates 
for the various Hubble type bins.  It is least well-defined for the elliptical 
and extreme late-type (Scd--Sm and I0) galaxies because of the small samples 
and limited dynamic range in velocity dispersions.  Thus, to first order, the 
gas velocities seem to track the random velocities of the stars.  
Quantitatively, however, the gas dispersions are systematically slightly {\it 
lower}\ than the stellar dispersions, on average by a constant offset of 
$\sim$20\%.  Defining the residuals from the \sigg\ = \sigs\ line by 
$\Delta \sigma \equiv  \log \sigma_g - \log \sigma_*$, 
$\langle \Delta \sigma \rangle  = -0.099\pm 0.010$.  
Barred and unbarred galaxies behave the same, with 
$\langle\Delta\sigma \rangle = -0.094\pm 0.014$ 
and $-0.104\pm 0.014$, 
respectively.  At the lowest velocities, $\sigma_*$ \lax\ 40 \kms, nearly all 
the objects have $\Delta \sigma > 0$.  This subgroup of high-$\Delta \sigma$ 
objects 

\vskip 0.3cm
\psfig{file=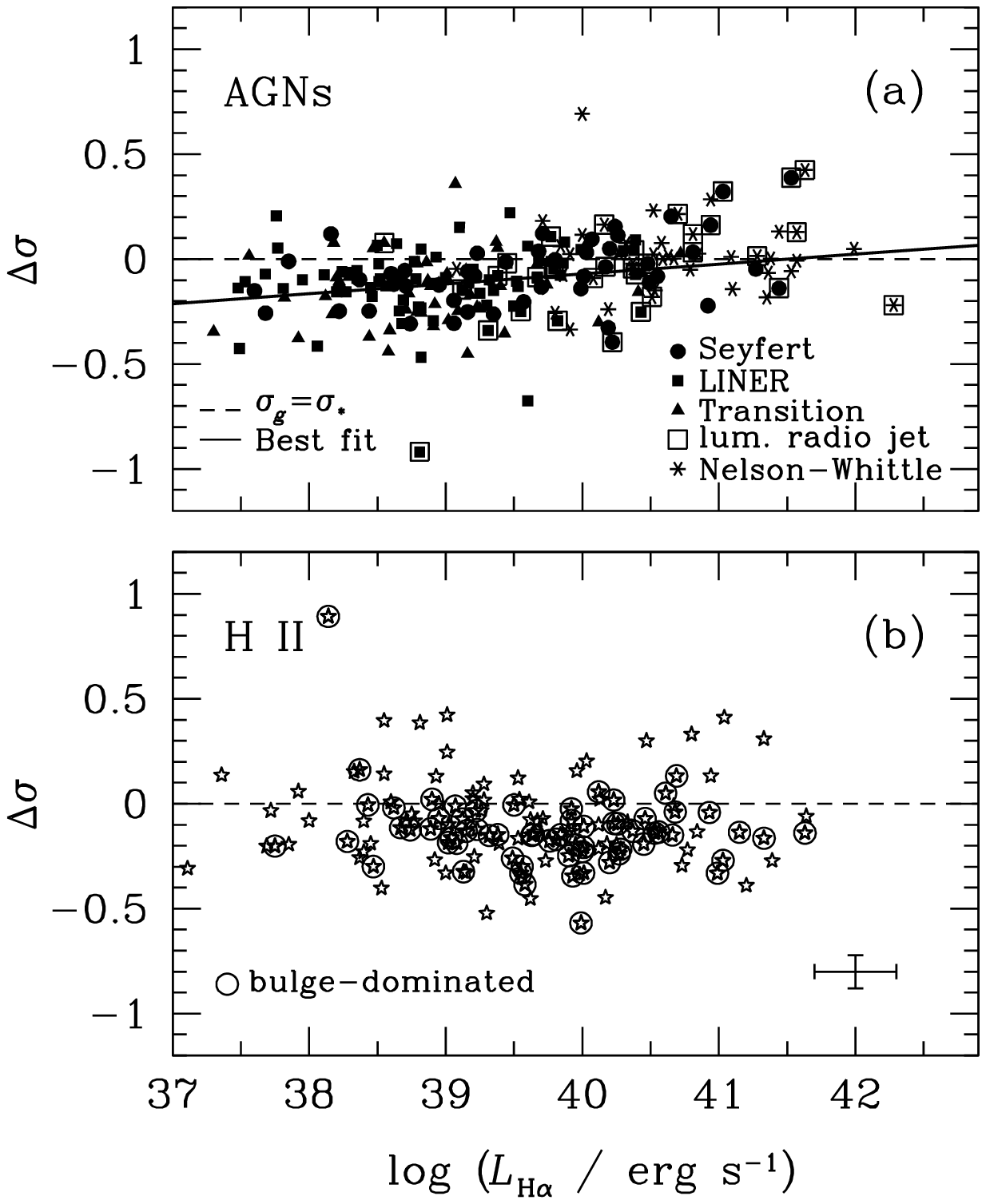,width=8.75cm,angle=0}
\figcaption[fig3.ps]{Residuals of the \sigg-\sigs\ correlation,
$\Delta \sigma \equiv \log \sigma_g - \log \sigma_*$, as a function of H\al\
luminosity \lhal, for ({\it a}) all AGNs (Seyferts, LINERs, and transition
objects), including the sample of luminous Seyferts from Nelson \& Whittle
(1995), and ({\it b}) \hii\ nuclei.  Objects with linear radio sources and
$P_{\rm 6 cm} > 10^{21.7}$ W~Hz$^{-1}$ are marked with boxes.  The dashed line
denotes \sigg\ = \sigs.  The solid line in panel ({\it a}) shows the best-fit
regression line.  A typical error bar is given in the lower-right corner of
the panel ({\it b}).
\label{fig3}}
\vskip 0.3cm

\noindent
cannot be attributed entirely to resolution 
effects because only a small number of \nii\ profiles are unresolved
and including them does not alter the distribution of data points.  Rather, it
appears that galaxy centers reach a ``floor'' at $\sigma_g \approx 30$ \kms.

In detail, the different Hubble types also have similar distributions:
$\langle \Delta \sigma \rangle =-0.15\pm 0.043$, 
$-0.105\pm 0.018$,    
$-0.113\pm 0.014$,    
and $-0.083\pm 0.026$ 
for E, S0--S0/a, Sa--Sbc, and Sc, 
respectively.  Ellipticals appear to have somewhat larger offsets than the 
disk galaxies, but the means of the different samples are not statistically 
different according to the Student's $t$ test.  By contrast, very late-type 
objects, in our sample dominated by Scd--Sm galaxies, clearly stand apart with 
$\langle \Delta \sigma \rangle = -0.036\pm 0.041$, 
differing mildly from the other groups at a significance of 94\%.  

\subsection{Dependence on Local Environment}

Ho et al. (1997a) give two parameters to gauge the tidal influence of nearby 
neighbors: (1) $\rho_{\rm gal}$, the density of galaxies brighter than $M_B = 
-16$ mag in the object's local vicinity, and (2) $\theta_p$, the projected 
angular separation to the nearest neighbor within a magnitude difference of 
$\pm$1.5 mag and a velocity difference of $\pm$500 \kms.  We find that 
$\Delta \sigma$ shows no correlation with either $\rho_{\rm gal}$ or $\theta_p$.

\subsection{Dependence on AGN Properties}

We next evaluate the dependence of the \sigg-\sigs\ correlation on nuclear
spectral classification (Fig.~2).  The four classes of emission-line objects
appear broadly similar, but in detail there are some subtle, important 
differences.  Taken at face value, the Seyfert sample shows the least degree 
of offset from the \sigg\ = \sigs\ line.  Its distribution of $\Delta \sigma$, 
with an average value of $-0.054\pm 0.025$, 
statistically differs from that of transition nuclei 
($\langle \Delta \sigma \rangle  = -0.13\pm 0.021$) 
at a significance level of 98\% according to the Student's $t$ test.  LINERs, 
with $\langle \Delta \sigma \rangle  = -0.095\pm 0.019$, 
lie sandwiched in between, although formally the differences are only 
statistically marginal.  Combining all three AGN subclasses, 
$\langle \Delta \sigma \rangle  = -0.096\pm 0.012$. 

As a group, \hii\ nuclei have an overall $\langle \Delta \sigma \rangle  
= -0.103\pm 0.017$.  
However, as mentioned in \S~3.1, late-type spirals, which largely dominate the 
\hii\ class, seem to comprise two distinct populations divided roughly at 
$\sigma_* \approx 40$ \kms.  Below this value, our sample of \hii\ nuclei has 
$\langle \Delta \sigma \rangle  = +0.28\pm 0.055$; 
above it, $\langle \Delta \sigma \rangle  = -0.15\pm 0.013$.  

Whittle (1992b, 1992c) and Nelson \& Whittle (1996) suggested that Seyfert 
galaxies with radio sources that are both strong {\it and}\ linearly extended 
tend to have emission lines that are systematically broader than the stellar 
velocities.  They interpret this to mean that the outflowing radio jets 
provide a secondary, nongravitational source of acceleration to the NLR gas.  
Whittle (1992b, 1992c) chose $P_{\rm 20 cm} = 10^{22.5}$ W~Hz$^{-1}$ as the 
threshold between strong and weak radio sources.  We have adopted this 
criterion\footnote{Whittle (1992b, 1992c) assumes a Hubble constant of $H_0$ = 
50 ${\rm km~s^{-1} Mpc^{-1}}$ and a reference wavelength of 20 cm, whereas we 
use $H_0$ = 75 ${\rm km~s^{-1} Mpc^{-1}}$ and a reference wavelength of 6~cm.  
The corresponding threshold, for a spectrum $f_\nu \propto \nu^{-0.7}$, is 
$P_{\rm 6 cm} = 10^{21.7}$ W~Hz$^{-1}$.}, and with the radio data for the 
Palomar low-luminosity AGNs presented in Ho \& Ulvestad (2001) and elsewhere 
(Ho 2008, and references therein), highlighted in Figure~2 the objects with 
luminous, jetlike radio sources.  Two of the objects in this subset---NGC~1068 
and NGC~3079---are clearly systematically offset above the rest.  But another 
six that satisfy the same radio criteria do not look extraordinary whatsoever, 
and not all high-$\Delta \sigma$ objects have strong radio jets.  The same 
trends hold for luminous radio sources among the LINERs and transition 
objects.  As a whole, strong radio jets are neither necessary nor sufficient 
to produce super-virial velocities in the NLR.

Closer inspection of Figure~2 reveals a more interesting effect within the AGN 
samples.  When the objects are flagged according to H\al\ luminosity, those 
with high luminosities---here, for illustrative purposes, chosen 
somewhat arbitrarily to be those above \lhal\ = $10^{40}$ \lum---tend to lie 
systematically displaced toward higher $\Delta \sigma$ compared to objects 
with lower luminosities. (We use the narrow component of the H\al\ luminosity, 
corrected for Galactic and internal extinction.) This is most clearly seen for 
the Seyfert sample.  It is less obvious for the LINERs and transition objects 
because of the paucity of luminous sources in these classes (Ho et al. 2003), 
but the same trend is noticeable nonetheless.  Within the Seyfert sample, 
$\langle \Delta \sigma \rangle  = +0.004\pm 0.043$ 
for objects with \lhal\ $\geq\,10^{40}$ \lum\ and 
$\langle \Delta \sigma \rangle  = -0.10\pm 0.026$  
for objects with \lhal\ $<\,10^{40}$ \lum.  The difference between the two 
means is significant at the level of 96\%.

To explore the luminosity effect further, we augmented the Palomar sample 
with 52 Seyfert nuclei from the work of Nelson \& Whittle (1995) that were not 
already in the Palomar survey.  These additional objects provide better 
coverage at the high-luminosity end.  H\al\ luminosities were converted from 
H\bet\ luminosities given in Nelson \& Whittle (1995) and Whittle (1992a), 
assuming H\al/H\bet\ = 3.1 (Gaskell \& Ferland 1984) and adjusted 
to a Hubble constant of $H_0$ = 75 ${\rm km~s^{-1} Mpc^{-1}}$.  Figure~3{\it a}\
plots $\Delta \sigma$ against \lhal\ for the combined sample of AGNs, again 
flagging the luminous, extended radio sources.  Note the nonzero 

\vskip 0.3cm
\psfig{file=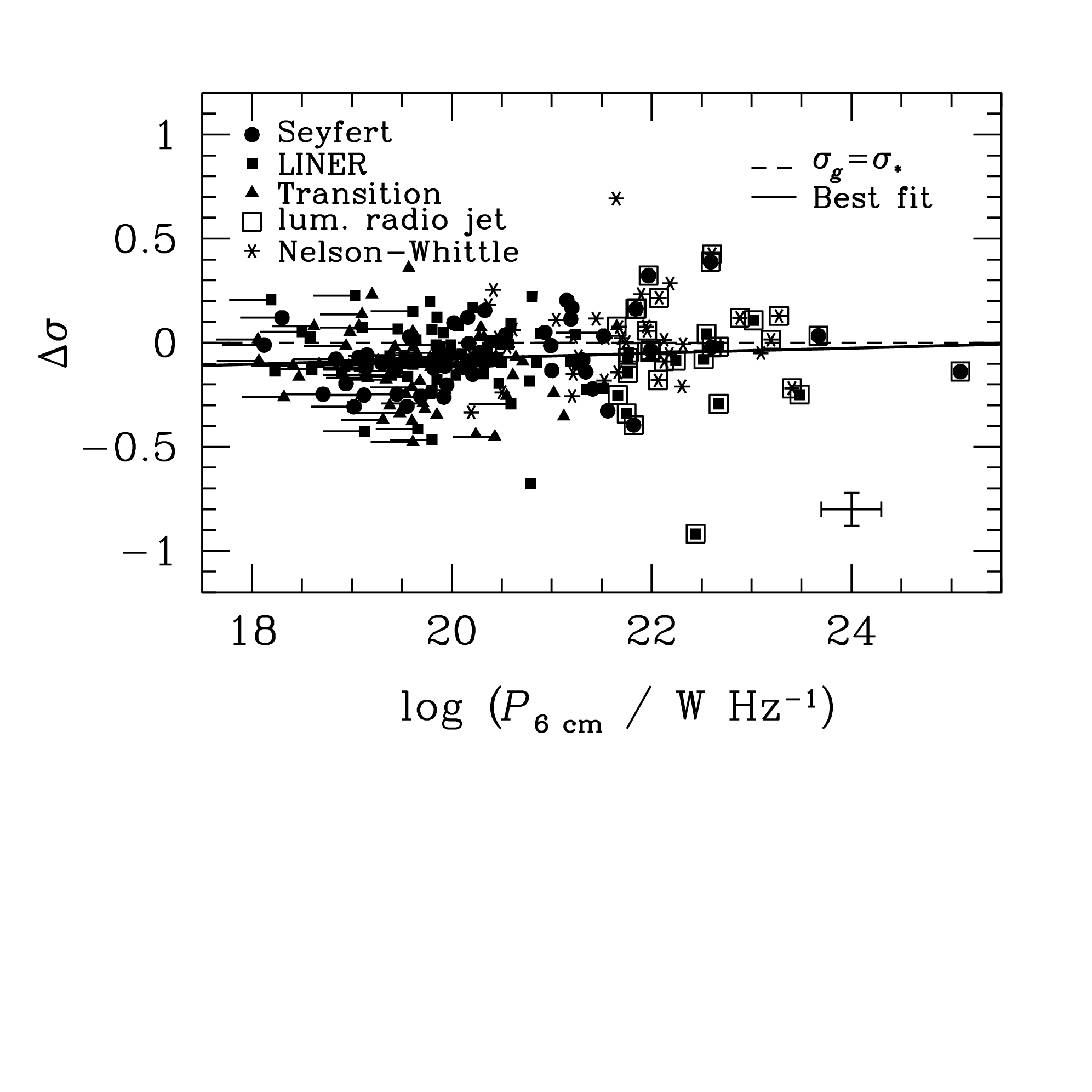,width=8.75cm,angle=0}
\figcaption[fig4.ps]{
Residuals of the \sigg-\sigs\ correlation for the AGNs (Seyferts, LINERs, and
transition objects in the Palomar survey, plus the luminous Seyferts from
Nelson \& Whittle 1995), $\Delta \sigma$, as a function of $P_{\rm 6 cm}$, the
6~cm power.  The symbols for the different sources are given in the figure
legend.  Objects with linear radio sources and $P_{\rm 6 cm} > 10^{21.7}$
W~Hz$^{-1}$ are marked with boxes.  The dashed line denotes \sigg\ = \sigs.
The solid line gives the best-fit regression line.  A typical error bar is
shown in the lower-right corner.
\label{fig4}}
\vskip 0.3cm

\noindent
slope.  A 
correlation can be seen for the Seyferts alone, as well as for the combined 
sample of LINERs and transition nuclei.  Treating log~\lhal\ as the 
independent variable, an ordinary least-squares regression fit to all the 
points gives

\begin{equation}
\Delta \sigma = (0.047 \pm 0.015) \log L_{{\rm H}\alpha} - (1.95 \pm 0.59),
\end{equation}

\noindent
with an rms scatter of 0.17 dex.   The scatter reduces to 0.16 dex with the 
luminous radio sources excluded.  The correlation has high statistical 
significance: the generalized Kendall's $\tau$ test yields $r = 0.47$ and 
$P_{\rm null} < 10^{-4}$.  Omitting the flagged radio sources has little 
effect on the results.  By contrast, the sample of \hii\ nuclei 
(Fig.~3{\it b}) does {\it not}\ show this effect, even though the \hii\ 
nuclei and AGNs span essentially the same range in H\al\ luminosities.  The 
different behavior of \hii\ nuclei compared to AGNs cannot be attributed to 
differences in morphological types.  Roughly half of the \hii\ nuclei are 
hosted in bulge-dominated (S0--Sbc; $T = -2$ to 4) galaxies, and these follow 
the same pattern as the rest of the sample (Fig.~3{\it b}).

In Figure~4, we substitute the H\al\ luminosity with the 6~cm radio power.  
Radio data for the Palomar Seyferts come from Ho \& Ulvestad (2001), and those 
for the supplementary sample of luminous objects come from Nelson \& Whittle 
(1995) and Whittle (1992a).  Data for the LINERs and transition objects are
taken from a variety of sources, as summarized in Ho (2008).  Again, we see a 
positive, significant trend, although formally it is weaker than that for 
$L_{{\rm H}\alpha}$.  In this case the Kendall's $\tau$ correlation 
coefficient is $r = 0.29$, with $P_{\rm null} = 6 \times 10^{-4}$.  A linear 
regression fit, calculated using the method of Schmitt (1985), which properly 
treats the few censored radio data points, gives 

\begin{equation}
\Delta \sigma = (0.013 \pm 0.0085) \log P_{{\rm 6cm}} - (0.34 \pm 0.17).
\end{equation}

\noindent
The tendency for $\Delta \sigma$ to increase with optical and radio power
suggests that the AGN luminosity, over a wide range of values, injects an 
additional, systematic contribution to the line broadening.  

Lastly, Figure~5 illustrates that an even stronger correlation exists between
$\Delta \sigma$ and the Eddington ratio, which is proportional to the mass 
accretion rate.  As in Greene \& Ho (2007), we estimate the bolometric 
luminosity from the H\al\ luminosity, converting $L_{{\rm H}\alpha}$ to 
$L_{{\rm 5100 \AA}}$ using the line-continuum correlation of Greene \& Ho 
(2005b) and then adopting $L_{\rm bol} = 9L_{{\rm 5100 \AA}}$.  

\vskip 0.3cm
\psfig{file=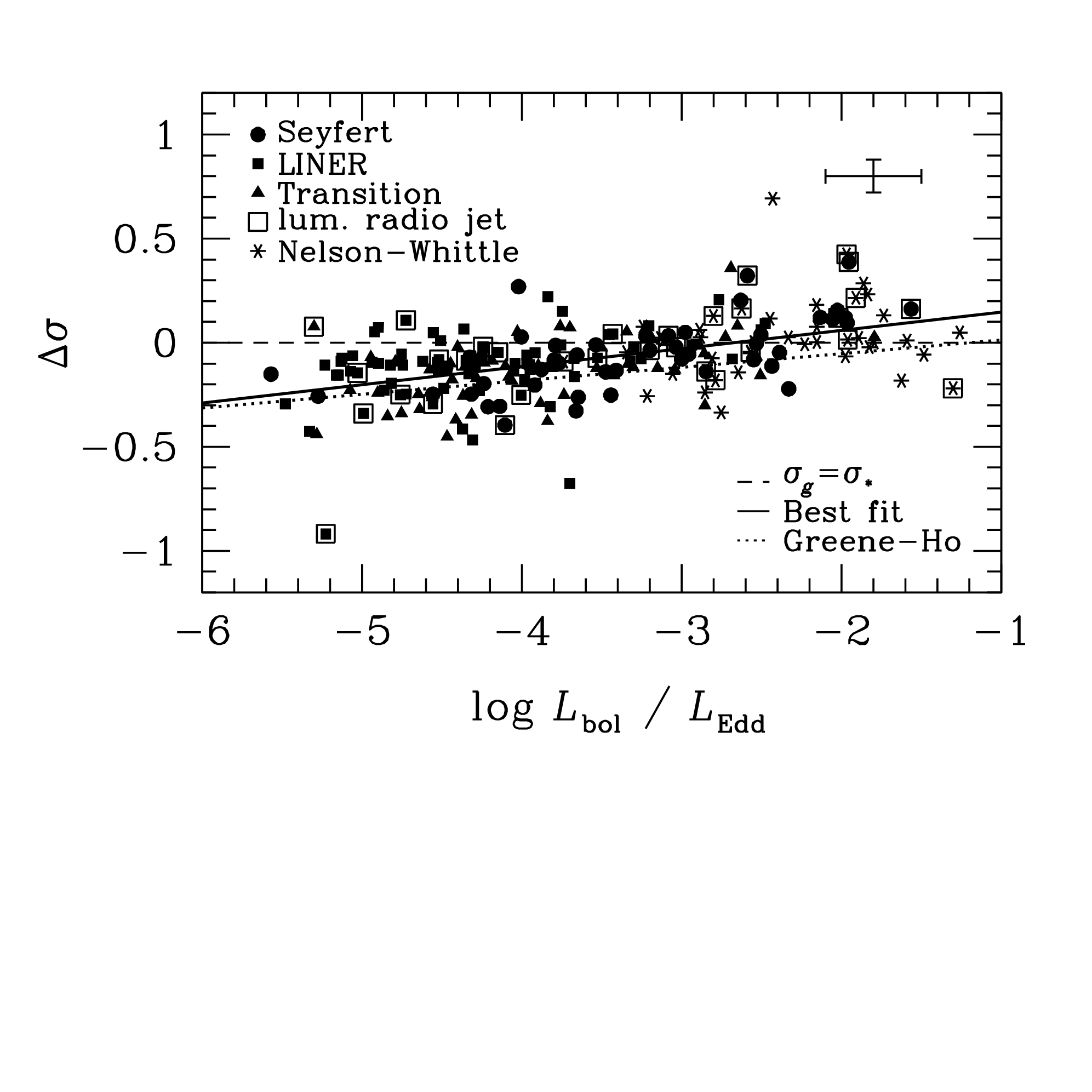,width=8.75cm,angle=0}
\figcaption[fig5.ps]{
Residuals of the \sigg-\sigs\ correlation for the AGNs (Seyferts, LINERs, and
transition objects in the Palomar survey, plus the luminous Seyferts from
Nelson \& Whittle 1995), $\Delta \sigma$, as a function of Eddington ratio,
\lbol/\ledd.  The symbols for the different sources are given in the figure
legend.  Objects with linear radio sources and $P_{\rm 6 cm} > 10^{21.7}$
W~Hz$^{-1}$ are marked with boxes.  The dashed line denotes \sigg\ = \sigs.
The solid line gives the best-fit regression line; the dotted line shows the
relation found by Greene \& Ho (2005a; Equation 4).  A typical error bar is
shown in the upper-right corner.
\label{fig5}}
\vskip 0.3cm

\noindent
We use the 
$M_{\rm BH}-\sigma_*$ relation of Tremaine et al. (2002) to obtain the black 
hole mass and hence the Eddington luminosity, $L_{\rm Edd}\,=\,
1.26\times10^{38} \left(M_{\rm BH}/M_{\odot}\right)$ \lum.  The Kendall's 
$\tau$ correlation coefficient now rises to $r = 0.74$, with $P_{\rm null} 
< 10^{-4}$.  An ordinary least-squares regression fit with 
log~$L_{\rm bol}/L_{\rm Edd}$ as the independent variable gives

\begin{equation}
\Delta \sigma = (0.087 \pm 0.011) \log L_{\rm bol}/L_{\rm Edd} + 
(0.23 \pm 0.042).
\end{equation}

\vskip 0.3cm
\noindent
The correlation is surprisingly tight, with an rms scatter of only 0.15 dex.
If we omit the sources with luminous radio jets, the scatter reduces even 
further to 0.14 dex.  $\Delta \sigma$ seems to correlate somewhat better 
with \lbol/\ledd\ than with \lhal, although formally, according to the $t$ and 
$F$ tests, neither the mean nor the variance of the residuals about the 
best-fit lines differs at a statistically significant level.

\subsection{Geometry and Kinematics of the Ionized Gas}

The gas line widths presented in this study were extracted from a single 
2\asec$\times$4\asec\ aperture centered on the nucleus, which corresponds to 
a region 200 pc $\times$ 400 pc for a typical distance of 20 Mpc (Ho et al. 
1997a).  Without spatially resolved data, it is impossible to know a priori in 
any individual object whether the line width represents true velocity 
dispersion or instead arises from spatial smearing of a steep inner rotational 
gradient.  Nevertheless, we can appeal to general arguments to test the 
scenario of rotational broadening.   If the nuclear gas is confined to a 
rotating disk aligned with the large-scale galactic plane, one might naively 
expect the line width to correlate with the galaxy inclination angle.  This 
effect is not seen in the Palomar survey (Ho 1996; Ho et al. 2003). However, 
as discussed by Whittle (1992b), this simple test is likely to be inconclusive 
because the line widths are primarily governed by virial velocities and AGN 
properties rather than by projection effects.  

Instead, Whittle (1992b), as did Heckman et al. (1989), examined the correlation
between inclination angle and the ratio $\Delta V_{\rm rot}$/FWHM, where 
$\Delta V_{\rm rot}$ is the observed rotation amplitude of the large-scale 
disk and FWHM is the full-width at half maximum of the central line profile.  
If the central nebular emission 

\vskip 0.3cm
\psfig{file=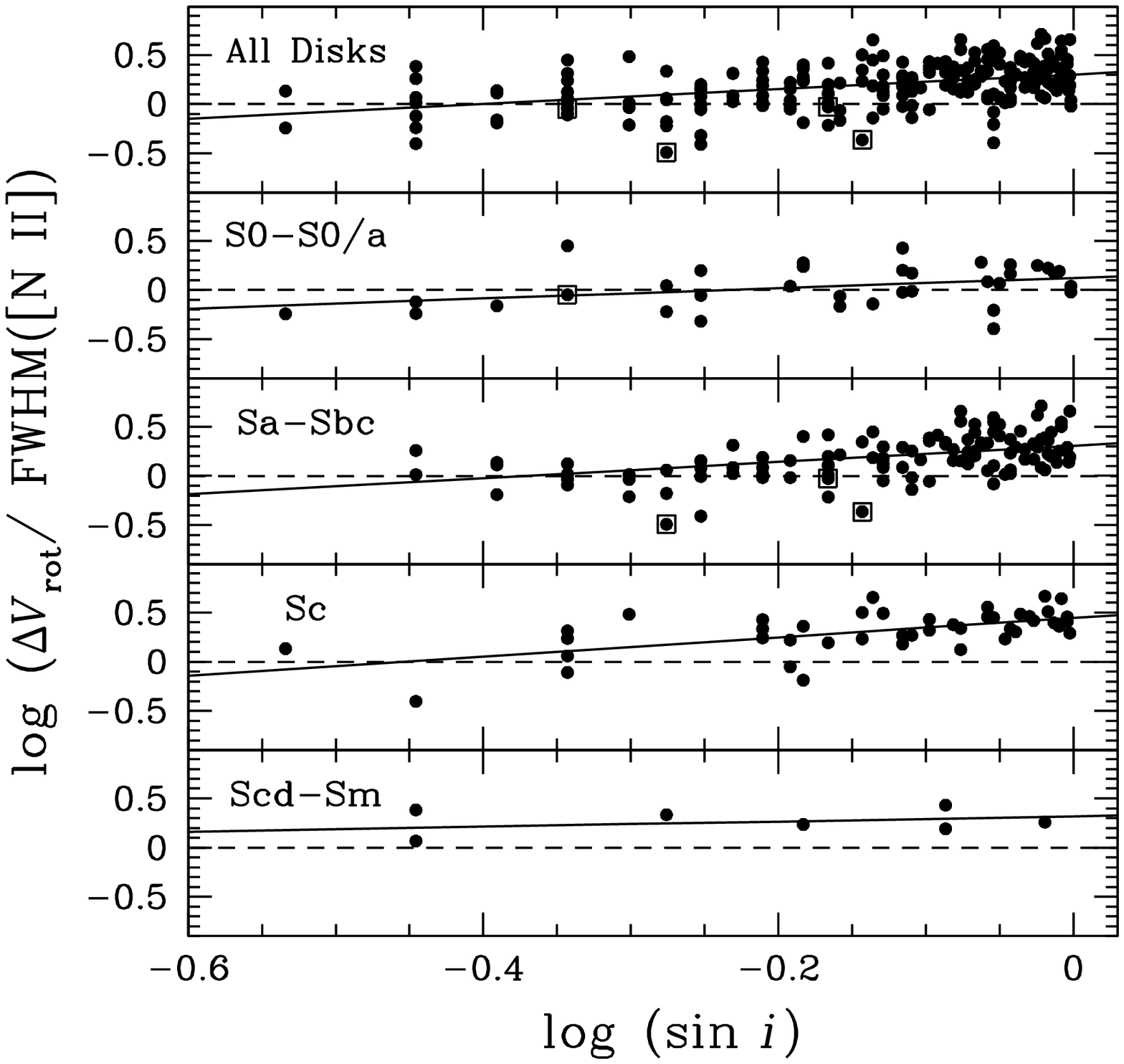,width=8.75cm,angle=0}
\figcaption[fig6.ps]{
Distribution of the ratio $\Delta V_{\rm rot}$/FWHM(\nii) versus sin~$i$
as a function of Hubble type.  The dashed line denotes an idealized model
wherein FWHM(\nii) arises entirely from rotation in the galactic plane.
Actual fits to the data, after excluding the luminous, linear
radio sources (marked with squares), are given as solid lines; the regressions
are calculated using an ordinary least-squares fit with $\log~{\rm sin}~i$ as
the independent variable.
\label{fig6}}
\vskip 0.3cm

\noindent
arises from a coplanar nuclear disk, FWHM, like 
$\Delta V_{\rm rot}$, is a projected quantity, and $\Delta V_{\rm rot}$/FWHM 
will be independent of inclination.  In the extreme alternative scenario where 
FWHM represents purely random velocity dispersion, $\Delta V_{\rm rot}$/FWHM 
should correlate positively with inclination.  The results of this exercise 
are shown in Figure~6 for all the non-elliptical galaxies, using data for 
$\Delta V_{\rm rot}$ (derived from integrated \hi\ line widths) and inclination 
angle $i$ from Ho et al. (1997a).  In each panel, the coplanar rotation model 
is marked with the horizontal dashed line.  The actual regression fit to 
the data (after excluding the flagged radio sources) is shown as a solid line.  
With the exception of the small number of very late-type (Scd--Sm) spirals, 
the vast majority of the sample strongly disagree with the coplanar rotation 
model.

As a final check, we also plotted the residuals $\Delta \sigma$ from Figure~1
against sin~$i$.  We find no correlation, as to be expected if \sigg\ and 
\sigs\ are both inclination-independent.

\vskip 1.5cm

\section{Discussion}

\subsection{Estimating \sigs\ from \sigg}

We combine new central stellar velocity dispersions (Ho et al. 2009) with
published emission-line width measurements (Ho et al. 1997a) to examine the 
relationship between the kinematics of the ionized gas and stars in the 
central regions of galaxies.  Unlike most studies that primarily focus on the
\oiii\ \lamb 5007 line, ours uses \nii\ \lamb 6583, which offers at least two 
major advantages.  \nii\ has a lower ionization potential and a lower critical 
density than \oiii, making it a better tracer over a larger radial extent of 
the bulge.  \nii\ also provides a more reliable probe of the gravitational 
potential because it is less susceptible to contamination by outflows or other 
radial motions.  Like most narrow emission lines in AGNs, \nii\ does exhibit 
line asymmetries indicative of radial flows (Ho 1996; Ho et al. 2003), but they 
seem to be less prevalent than in high-ionization transitions such as \oiii\ 
(Heckman et al. 1981; Vrtilek \& Carleton 1985; Whittle 1985).  Studies of 
AGN line profiles that include transitions from a wide range of ionization 
states (Busko \& Steiner 1992; Greene \& Ho 2005a) find that \sii\ \lamb\lamb
6716, 6731 generally tend to be less asymmetric than \oiii\ (but see Rice 
et al. 2006).  Since the profile of \sii\ empirically matches well the profile 
of \nii\ (e.g., Ho et al. 1997d), it follows that \nii\ is also less 
asymmetric than \oiii.  Moreover, the average line centroid of \nii\ is 
consistent with the systemic velocity of the host galaxy, whereas 
high-ionization lines such as \oiii\ are often blueshifted (Boroson 2005).  
Previous work that makes use of \nii\ to probe the kinematics of the NLR 
include those of Phillips et al. (1986), Verdoes~Kleijn et al. (2006), Zhou et 
al. (2006), Chen et al. (2008), and Walsh et al. (2008).

Our principal finding is that, within the central few hundred parsecs of 
galaxy bulges, the velocity dispersion of the ionized gas is comparable to, 
but typically somewhat lower than, the velocity dispersion of the stars.  The 
rough correspondence between \sigg\ and \sigs\ indicates that
the NLR is in dynamical equilibrium with the stellar potential.  Defining 
$\Delta \sigma = \log \sigma_g - \log \sigma_*$, $\Delta \sigma$
ranges from $-0.15$ to $+0.08$, with an average value of $-0.099$, or $\langle 
\sigma_g/\sigma_* \rangle = 0.80$.  In detail, $\Delta \sigma$ shows, at best, 
a mild variation with galaxy morphology but no discernible dependence on the 
presence of a large-scale bar or local galaxy environment.  By contrast, the 
more luminous AGNs studied by Whittle (1992b) seem to exhibit broader lines in 
barred and interacting galaxies.  

As an aside, we note that the late-type spirals in our sample show an 
intriguing property: below $\sigma_* \approx 40$ \kms, $\Delta \sigma$ is 
{\it always}\ positive.  It seems that in these late-type galaxies with small 
bulges (possibly all ``pseudo-bulges''; Kormendy \& Kennicutt 2004), the 
velocity dispersion of the warm ($\sim10^4$ K) ionized gas has a floor of 
$\sim 30$ \kms.  This might represent a minimum threshold of interstellar 
turbulence in the central regions of present-day disk-dominated galaxies, 
maintained, perhaps, through stellar feedback associated with their modest 
star formation rates (Ho et al. 1997c).  The dispersion threshold we find is
qualitatively similar to the magnitude of non-ordered motions detected in the 
central regions of low-surface brightness galaxies (e.g., Kuzio~de~Naray et 
al. 2008; Pizzella et al. 2008a, 2008b).

The availability of a uniform set of nuclear spectroscopic classifications for 
the Palomar galaxies has prompted us to reexamine the kinematics of the NLR, 
their relationship with the stellar potential of the bulge of the host galaxy, 
and the possible role of nongravitational perturbations by AGNs.  Previous 
treatment of this subject has concentrated almost exclusively on radio galaxies
(Smith et al. 1990; Verdoes~Kleijn et al. 2006; Noel-Storr et al. 2007) and 
relatively luminous Seyfert nuclei (V\'eron 1981; Terlevich et al. 1990; 
Veilleux 1991; Whittle 1992b, 1992c; Nelson \& Whittle 1996; Jim\'enez-Benito 
et al. 2000; Botte et al. 2005; Greene \& Ho 2005a; Bian et al. 2006; Zhou et 
al. 2006; Chen et al. 2008; Dasyra et al.  2008).  The reviews by Wilson \& 
Heckman (1985) and Whittle (1993) included some preliminary results on LINERs, 
although they were based on rather fragmentary data.  They concluded that the 
NLR kinematics of LINERs, as in Seyferts, are largely virial.  In their survey 
of ionized gas in E and S0 galaxies, most of which contain LINER nuclei, 
Phillips et al. (1986) compared the FWHM of \nii\ \lamb 6583 with the absolute 
magnitude of the host galaxies and inferred that in general \sigg\ \lax\ \sigs.

Our analysis strongly reinforces the notion that gravity dominates the 
kinematics of the NLR, not only in Seyferts, but indeed in {\it all}\ 
emission-line nuclei found in nearby galaxies, including low-ionization, 
low-luminosity AGNs such as LINERs and transition objects, and even \hii\ 
nuclei.  The gas velocity dispersions are comparable to, but typically 
somewhat less than, the stellar velocity dispersions measured over a
physical scale of typically $200-400$ pc.  The velocity 
discrepancy observed in \hii\ nuclei and AGNs is about the same, 
$\langle$\sigg/\sigs$\rangle \approx 0.8$, with a surprisingly small standard 
deviation of only $\sim 0.2$ dex.  As in the work of Whittle (1992b, 1992c) 
and Nelson \& Whittle (1996), we find that powerful radio jets, in the 
minority of nearby sources where they are found, do sometimes seem to exert 
secondary, super-virial acceleration on the NLR gas, although the 
correspondence is hardly one-to-one.  However, the main new insight obtained 
from our work is that the impact of the central AGN may be more widespread and 
systemic than previously suspected.  We find that, not only among Seyfert 
galaxies but also in the characteristically less powerful LINERs and 
transition nuclei, the offset between \sigg\ and \sigs\ varies mildly but 
{\it systematically}\ with the level of AGN activity.  The effect can be seen 
with nuclear activity as gauged through optical (H\al) luminosity 
(Fig.~3{\it a}), radio luminosity (Fig.~4), and especially Eddington ratio
(Fig.~5).  The weakest AGNs in the Palomar sample (LINERs and transition 
nuclei with \lhal\ $\approx 3\times10^{37}$ \lum, $P_{\rm 6cm}$ \lax\ $10^{18}$ 
W~Hz$^{-1}$, and \lbol/\ledd\ $\approx 3\times 10^{-6}$) are characterized by
\sigg/\sigs$\approx 0.6-0.8$, while luminous sources (Seyferts with \lhal\ 
$\approx 3\times10^{41}$ \lum, $P_{\rm 6cm} \approx 10^{23}$ W~Hz$^{-1}$, and 
\lbol/\ledd\ $\approx 4\times 10^{-2}$) can attain \sigg/\sigs$\approx 1.2-1.4$.
Remarkably, \hii\ nuclei, even those spanning a similar range in H\al\ 
luminosity and Hubble type as the AGNs, do {\it not}\ follow these trends 
(Fig.~3{\it b}).  Evidently the physical nature of the activity---stellar or 
nonstellar---matters.

It is worth noting that Whittle (1992b, 1992c) and Nelson \& Whittle (1996)
find $\langle$\sigg/\sigs$\rangle \approx 0.9$ for their sample of Seyferts 
(after excluding the luminous, jetlike radio sources), somewhat larger than 
the average value of $\langle$\sigg/\sigs$\rangle \approx 0.8$ for the Palomar 
objects.   This difference is easy to explain in view of the luminosity 
dependence just mentioned, since Whittle's sample is more luminous than the 
Palomar objects.  In their analysis of emission-line objects from the Sloan 
Digital Sky Survey (SDSS), Chen et al. (2008) find, as we do in this study, 
that \sigg/\sigs\ varies slightly but systematically with nuclear spectral 
classification: transition objects have the lowest values and Seyferts the 
highest, with LINERs in between.  This trend reflects none other than the 
dependence of \sigg/\sigs\ on luminosity or Eddington ratio.  Ho (2008, 2009) 
proposes that the mass accretion rate, which scales with the Eddington ratio, 
is the primary physical driver responsible for variations in spectral 
classification.  The accretion rate increases systematically along the 
sequence transition objects $\rightarrow$ LINERs $\rightarrow$ Seyferts.

If the size of the NLR scales with luminosity, Laor (2003) suggests that the
NLR in low-luminosity AGNs may be sufficiently compact that its kinematics may
be influenced more by the gravitational potential of the central black hole
than by that of the surrounding bulge.  Spatially resolved observations of the
NLR in some LINERs at {\it Hubble Space Telescope}\ resolution indeed do find
that the line widths are largest interior to the black hole's gravitational
sphere of influence, gradually merging at larger radii with the stellar
dispersion of the bulge (Walsh et al. 2008).  If this effect were dominant, 
however, \sigg/\sigs\ would rise with decreasing luminosity, exactly the 
opposite of what is seen in our sample.  Evidently on a scale of several 
hundred parsecs the integrated kinematics of the NLR are controlled primarily
by the gravitational potential of the stars and by AGN feedback.

The discovery of a tight relation between central black hole mass and bulge
stellar velocity dispersion (the \mbh-\sigs\ relation: Gebhardt et al. 2000;
Ferrarese \& Merritt 2000) has prompted renewed interest in efficient methods
to estimate \sigs, especially in objects such as luminous AGNs where this
parameter is challenging to measure directly (Greene \& Ho 2006).  In view of 
the approximate equality between \sigg\ and \sigs\ and the ease with which 
\oiii\ \lamb5007 can be detected in AGNs, Nelson (2000) suggested that the 
stellar velocity dispersion of the host's bulge can be estimated by assuming
\sigs\ = \sigg(\oiii).  Numerous studies have since adopted this strategy to 
investigate the \mbh-\sigs\ relation in AGNs (e.g., Wang \& Lu 2001; Grupe \& 
Mathur 2004) and its possible evolution with redshift (e.g., Shields et al. 
2003; Salviander et al. 2007).  However, the large observed scatter of the 
\sigg-\sigs\ correlation (Boroson 2003), not to mention the kinematic 
anomalies long known to afflict the \oiii\ line (Boroson 2005), gave cause for 
concern.  To mitigate these uncertainties, Greene \& Ho (2005a) recommended 
that only the core of the \oiii\ line be used, and, if possible, that \oiii\ 
be abandoned altogether in favor of lower-ionization transitions such as \oii\ 
\lamb3727 or \sii\ \lamb\lamb6716, 6731.

This study shows that the width of \nii, another low-ionization line, also 
traces the stellar velocity dispersion of the bulge.  The scatter is modest,
$\sim 0.2$ dex, but not insignificant.  Since \mbh\ $\propto$ \sigs$^4$ 
(Tremaine et al. 2002), a 0.2 dex uncertainty on \sigs\ translates into a 
0.8 dex (or factor of $\sim$6) uncertainty on \mbh.  Importantly, we find that 
the velocity dispersion of the ionized gas does not identically match the 
velocity dispersion of the stars, but rather, on average, \sigg/\sigs$\approx 
0.8$.  Closer examination reveals that the residuals of the \sigg-\sigs\ 
relation in fact depend on AGN properties.  The principal ``second parameter'' 
appears to be either the Eddington ratio (\lbol/\ledd) or some measure of the 
AGN luminosity (we use \lhal).  The residuals seem to correlate somewhat 
better with \lbol/\ledd\ than with \lhal, although from our data alone we 
cannot truly tell which variable is more fundamental.  Since the optical NLR 
luminosity correlates strongly with radio continuum luminosity (e.g., Ho \& 
Peng 2001; Ulvestad \& Ho 2001; Nagar et al. 2005), it is hardly surprising 
that $\Delta \sigma$ correlates with radio power too, although the scatter is 
larger and the statistical significance is inferior compared to either \lhal\ 
or \lbol/\ledd.  The dependence of the residuals of the \sigg-\sigs\ relation 
on \lbol/\ledd\ was first noticed by Greene \& Ho (2005a; later confirmed by 
Bian et al. 2006) for a large sample of more luminous AGNs selected from SDSS, 
and, within the uncertainties, the Palomar sources seem to follow roughly the 
low-luminosity extrapolation of the SDSS sample (Fig.~5).  Curiously, Greene 
\& Ho did not see any correlation between $\Delta \sigma$ and AGN optical 
(\oiii) luminosity or radio power.  It is possible that the large linear 
scales probed by SDSS (the 3\asec\ fibers correspond to 5.4 kpc at $z = 0.1$) 
introduce substantial uncertainties into measurements of the bulge and NLR.

The dependence of $\Delta \sigma$ on AGN properties can be exploited to sharpen
the \sigg-\sigs\ correlation as a tool to predict \sigs\ when the latter is, 
as is often the case, difficult or impractical to measure directly in AGNs.  
Even when a low-ionization line or only the core of the \oiii\ line is being 
used, as recommended by Greene \& Ho (2005a), it is important to realize that 
\sigg/\sigs\ is {\it not}\ identically equal to unity.  Rather, \sigg/\sigs\ 
varies from $\sim 0.6$ to $\sim 1.4$ as a function of luminosity or Eddington 
ratio.  Equations (1) and/or (3) should be used to obtain a more refined 
estimate of \sigs. Moreover, significant scatter still remains ($\sim 0.15$ 
dex), even after correcting for the second parameter, most of which may be 
intrinsic and irreducible owing to the complexities of the NLR (Rice et al. 
2006).

\subsection{Evidence for AGN Feedback}

From a physical point of view, what is the origin of the $\Delta \sigma$-\lhal\
or $\Delta \sigma$-\lbol/\ledd\ correlations?  As discussed further below, 
if the gas derives principally from mass loss from bulge stars, its kinematics
should generally track the kinematics of the stars.  But because the 
gas is collisional and experiences hydrodynamical drag against the surrounding 
hot medium, we expect it to be kinematically slightly colder than the stars.
In the absence of additional energy input from other sources, we anticipate 
\sigg/\sigs\ \lax\ 1, as observed.  As additional energy is injected into the 
system, for example from activation of the central black hole, the gas gains 
energy, to the point that \sigg\ approaches or even overtakes \sigs.  Precisely
how this is accomplished is unclear.  Do AGNs impart mostly mechanical energy 
from outflows/winds or more highly collimated fast jets?  Does radiation 
pressure matter?  The empirical trends presented in this paper offer some 
insights into these issues, which ultimately have bearing on current interests 
in AGN feedback in galaxies.

The fact that the emission-line widths become systematically broader with 
increasing AGN activity---whatever the principal source of energy---implies 
that the gas is primarily accelerated rather than directly heated.  The H\al\ 
luminosity, and presumably also the bolometric luminosity, strongly correlates 
with $\Delta \sigma$, but it seems doubtful that the radiation density itself
plays a central role because \hii\ nuclei with the same range in luminosity 
and sampled over a similar volume do not show the same effect.  Compared to 
nuclear star formation, AGNs have a characteristically harder ionizing 
spectrum and a more centrally concentrated radiation field.  Whether these 
factors result in more efficient radiative acceleration of the ionized gas in 
AGNs needs to be further investigated.  

The radio-emitting plasma is an obvious suspect---that is, a scaled-down, but 
qualitatively similar agent as that responsible for super-virial acceleration 
in the most extreme, jet-dominated sources.  This is a tenable hypothesis 
because nearly all Seyferts (e.g., Kukula et al. 1995; Ho \& Ulvestad 2001; 
Ulvestad \& Ho 2001) and LINERs (Nagar et al. 2005) and a sizable fraction of 
transition objects (Filho et al. 2000, 2002) contain nuclear radio sources, 
many with morphological structures resembling jets.  The radio luminosity 
function of low-luminosity AGNs varies continuously over at least 4 orders of 
magnitude in radio power, from $P_{\rm 20 cm}\,\approx\, 10^{19}$ to 
$10^{23}$ W~Hz$^{-1}$ (Ulvestad \& Ho 2001; Filho et al. 2006).  There is no 
characteristic power that might demarcate a luminosity threshold above which 
radio sources become effective in perturbing the NLR.   Although the degree of 
radio-loudness (expressed as the relative fraction of the radiative output 
emerging in the radio band) actually decreases with increasing Eddington ratio 
(Ho 2002; Terashima \& Wilson 2003; Greene et al. 2006), Ho (2008) notes that 
extended, jetlike structures are more prevalent in high-\lbol/\ledd\ sources 
(Seyferts) than in low-\lbol/\ledd\ sources (LINERs).  The radio cores in 
LINERs, despite being energetically more dominant, may be less effective at 
stirring up the NLR because of their compact structure.  The extended, linear 
features in Seyferts, on the other hand, can impact a larger pool of the 
circumnuclear material.  Jet-induced interactions have been invoked to explain 
the morphology and kinematics of the NLR in a number of well-studied Seyfert 
galaxies (e.g., Mrk~78: Whittle \& Wilson 2004; NGC~1068: Das et al. 2006; 
NGC~4151: Mundell et al. 2003).  Despite the natural appeal of radio jets, 
however, our analysis shows that radio power correlates more poorly with 
$\Delta \sigma$ than H\al\ luminosity or Eddington ratio.

We speculate that the dominant source of energy input comes from accretion disk
winds, a common feature in AGNs (Crenshaw et al. 2003).  Disk winds or outflows
presumably agitate the NLR through shocks, provided that they do not violate 
other constraints that limit the role of shocks in the excitation of the gas 
(Ho 2008).  The ability of an accretion disk to generate a wind may depend 
critically on \lbol/\ledd\ (Proga 2007), and systems such as LINERs with 
ultra-low \lbol/\ledd\ seem to lack outflows altogether (Ho 2008).  
Consistent with Greene \& Ho (2005a), we propose that the Eddington ratio 
primarily accounts for the secondary line broadening in the NLR.  The apparent 
interchangeability between \lbol/\ledd\ and \lhal\ is probably an artifact of 
the small dynamic range in black hole mass for the majority of the Palomar 
sample, coupled with the sizable intrinsic scatter in the \sigg-\sigs\ relation.

\subsection{Line Width-Luminosity Correlation in AGNs}

Phillips et al. (1983) first noticed that the luminosity of the \oiii\ line
in Seyfert galaxies loosely correlates with its line width.  This result has
since been extended to much larger samples of Seyferts by Whittle (1985, 1992b)
and Gu et al. (2006), as well as to lower-luminosity AGNs in the Palomar survey
by Ho et al. (2003; using H\al\ for line luminosity and \nii\ for line width).
The form of the correlation is $L_{\rm NLR} \propto {\rm FWHM}^a$, where,
depending on the study and sample, the slope can take a wide range of values,
from $a \approx 3$ to 6, with a typical value of $\sim 4$.  The physical
origin of the line width-luminosity relation has been unclear.  Whittle
(1992b) suggests that the line width-luminosity relation is actually a
secondary correlation, reflecting on the one hand the more primary link
between bulge mass and NLR velocity, and on the other various possible
interdependences among bulge mass, NLR luminosity, radio power, and line
width.

In light of the results of this study and other recent developments, we offer
a simple explanation.  We have confirmed, as has long been suspected, that
the widths of the narrow emission lines largely reflect the gravitational
potential of the bulge, such that FWHM $\propto$ \sigs.  The bulge, meanwhile,
directly links to the mass of the central black hole via the fundamental
relation $M_{\rm BH} \propto \sigma_*^4$ (Tremaine et al. 2002).  Now, a given
$M_{\rm BH}$ radiates up its Eddington luminosity, which in turn is related to
the NLR luminosity by a bolometric correction and the Eddington ratio
(accretion rate) of the system.  Hence, AGNs should populate a distribution
whose {\it upper envelope}\ follows a ridgeline defined by $L_{\rm NLR} \propto
{\rm FWHM}^4$.  At a given FWHM, $L_{\rm NLR}$ has a maximum but no hard
minimum (other than that dictated by detection sensitivity).  Indeed, this is
what is seen.  Ho et al. (2003) presented the line width-luminosity correlation
separately for each of the three classes of AGNs in the Palomar survey, and
it is clear that the zero point of the correlation shifts with the average
luminosity of each class.

\subsection{Source of the Gas and Its Dynamical Support} 

As discussed in Ho (2008, 2009), the gas that ultimately fuels AGNs in nearby 
galaxies can be readily supplied internally by mass loss from evolved stars 
(red giants and planetary nebulae) in the bulge.  The same argument, the gist 
of which was already advanced by Minkowski \& Osterbrock (1959) to explain the 
origin of nebular emission seen in ellipticals, can be extended to include the 
entire gas reservoir that makes up the NLR in low-luminosity AGNs.  Assuming 
$T_e = 10^4$ K, $n_e = 100$ \cc, H\al/H\bet\ = 3.1, and an effective 
recombination rate for H\al\ of $\alpha_{\rm H\alpha}^{\rm eff} = 
9.36\times10^{-24}$ cm$^3$~s$^{-1}$ (Osterbrock 1989), 

\begin{equation}
M_{\rm NLR} = 2.97 \times 10^3 \left(\frac{100~{\rm cm}^{-3}}{n_e}\right) 
\left(\frac{L_{{\rm H}\alpha}}{10^{38}~{\rm erg}~{\rm s}^{-1}}\right) 
\,\,\, M_{\odot}.
\end{equation}

\noindent
From the population statistics of the Palomar survey assembled in Ho et al. 
(2003), LINERs and transition objects have median \lhal\ $\approx\, 7\times 
10^{38}$  \lum\ and $n_e \approx 200$ \cc, and hence $M_{\rm NLR} \approx 
1\times 10^4$ \solmass.  The corresponding values for Seyferts are \lhal\ 
$\approx\, 40 \times 10^{38}$ \lum, $n_e \approx 400$ \cc, and $M_{\rm NLR} 
\approx 3\times 10^4$ \solmass.  Strictly speaking, these values only pertain 
to the central 200 pc $\times$ 400 pc region sampled by the spectroscopic 
aperture, but given the high degree of central concentration of the 
line-emitting gas (Ho 2008), they probably underestimate the integrated 
values for the entire NLR by no more than a factor of $\sim 2$.  These mass 
estimates can be compared with the amount of material shed from evolved stars.
For a Salpeter stellar initial mass function with a lower-mass cutoff of 0.1 
\solmass, an upper-mass cutoff of 100 \solmass, solar metallicities, and an 
age of 15 Gyr  (Padovani \& Matteucci 1993), 

\begin{equation}
\dot M_*\,\approx\,3\times10^{-11} \, 
\left(\frac{L_*}{L_{\odot, V}}\right) \,\,\, M_{\odot}\,{\rm yr}^{-1}.
\end{equation}

\noindent
With a median Hubble type of Sa, the host galaxies of the Palomar AGNs have 
sizable bulges and correspondingly healthy gas supply rates.  For a median 
bulge luminosity of $M_B \approx -19.3$ mag (Ho et al. 2003) and $B-V = 0.8$ 
mag (Fukugita et al. 1995), $\dot M_* \approx 0.5\, M_{\odot}\, {\rm yr}^{-1}$.
These values pertain to the entire bulge.  Within the spectroscopic aperture 
the nuclear stellar continuum luminosities (Ho et al. 1997a) are approximately 
a factor of 10 lower than the integrated luminosities, which implies $\dot M_* 
\approx 0.05\, M_{\odot}\, {\rm yr}^{-1}$. Thus, stellar mass loss can sustain 
the gas reservoir in the NLR if the stellar debris survives for periods longer 
than $\sim 2\times10^5-10^6$ yr before it dissipates and merges with the 
surrounding hot interstellar medium (Mathews 1990; Parriott \& Bregman 2008).

If the ionized gas is mainly produced internally through stellar mass loss
in the bulge\footnote{We do not mean to suggest that external acquisition of 
gas does not take place.  Indeed, an external origin has been invoked to 
explain the distribution of kinematic axis misalignments between the gas and 
stars (Bertola et al. 1992; Caon et al. 2000; Sarzi et al. 2006), as well as 
certain H~I properties (Ho 2007), in early-type galaxies.}, what is the 
expected kinematical signature of this material?  The supersonic motion of the 
ejecta through the ambient medium generates a strong shock that thermalizes 
the gas to the kinetic temperature of the stars, $\sim 10^6-10^7$ K, on a 
timescale of $\sim 10^5-10^6$ yr (Sanders 1981; Mathews 1990).  Not all of the 
gas gets heated, however.  Numerical simulations of mass loss from evolved 
stars in elliptical galaxies, after accounting for radiative cooling, predict 
that some fraction of the gas, perhaps $\sim 20\%$, remains cool (Parriott \& 
Bregman 2008).  Photoionization of this residual gas, as well as the fresh 
ejecta prior to thermalization, produces the nebular emission visible in the 
optical.\footnote{The optical line emission in ellipticals and bulges in 
principle can be excited by a variety of mechanisms, including photoionization 
by young or old hot stars, photoionization by a central AGN, shocks, 
self-irradiation from cooling condensations, and even cosmic ray heating by 
the radio-emitting plasma.  With the possible exception of the weakest 
emission-line objects, however, Ho (2008) argues that the majority of the 
low-luminosity AGNs in the Palomar survey---essentially all nearby galaxies 
with sizable bulges---are powered by nonstellar photoionization.}  During the 
early stages of their evolution, the mass loss envelopes orbit roughly 
cospatially with their parent stars and thus closely track their natal 
velocities.  As bulge stars do not have a strong rotational component to their 
velocities, neither should the gas shed by them.  After normalizing the gas 
line widths by the observed (projected) rotational velocity of the large-scale 
(\hi) disk (Fig.~6), the significant, positive trend with galaxy inclination 
angle that remains implies that the ionized gas does not reside in a plane 
aligned with the large-scale galaxy.  Moreover, the ejecta quickly 
lose memory of their birth sites as the material experiences drag deceleration 
against the surrounding hot medium.  As discussed by Mathews (1990), the warm, 
optical-emitting clouds cannot be self-gravitating.  Instead, they must be 
pressure-confined by the ambient hot medium, and hence their kinematics will 
largely imprint the local velocity of the hot gas.  Since the hot gas is in 
virial equilibrium with the bulge stars, to first order we anticipate 
$\sigma_g \approx \sigma_*$.  In detail, our analysis finds $\sigma_g \approx 
0.8 \sigma_*$.  As the referee points out, supersonic turbulence in the hot 
gas rapidly dissipates through shocks.  It is thus natural for $\sigma_g$ to 
be slightly subvirial because the turbulence---the source of the random 
motions---is subsonic.  When an AGN is present, the hot gas may be more 
agitated.

The scenario presented above provides a useful framework for interpreting a 
number of trends gleaned from spatially resolved observations of the central 
regions of galaxies.  {\it Hubble Space Telescope}\ images frequently reveal 
dust lanes, spiral 
patterns, and a variety of other small-scale fine features in galaxy centers, 
but rarely do the structures ever take the form of thin, well-defined disks 
(Phillips et al. 1996; Carollo et al. 1997; Malkan et al. 1998; Pogge et al. 
2000; Ho et al. 2002; Martini et al. 2003; Sim\~{o}es Lopes et al. 2007).  
Since the dust generally traces the nebular gas (e.g., Ravindranath et al. 
2001; Falc\'on-Barroso et al. 2006), we can infer that the ionized gas more or 
less follows the same topology as the dust.  The complex morphology of the 
gas strongly hints that its velocity field must be similarly chaotic.
Where available, observations support this picture.  For example, a number of 
authors have noted that the central rotation curves of the ionized gas often 
rise more slowly than the circular velocities predicted from the luminosity 
profile of the stars (Caldwell 1984; Caldwell et al. 1986; Fillmore et al. 
1986; Kent 1988; Kormendy \& Westpfahl 1989; Cinzano \& van~der~Marel 
1994; Bertola et al. 1995; Fisher 1997; Cinzano et al. 1999; Pignatelli et al. 
2001), suggesting that random motions contribute appreciably to the dynamical 
support of the gas.  On scales probed by the {\it Hubble Space Telescope}, 
very few objects exhibit clean signatures of dynamically cold disks undergoing 
circular rotation (Sarzi et al. 2001; Ho et al. 2002; Atkinson et al. 2005; 
Noel-Storr et al. 2007).  Although the statistics are still quite limited, 
evidence is emerging that the degree of kinematic disturbance may be directly 
linked to the level of AGN activity (Verdoes~Kleijn et al. 2006; Dumas et al. 
2007; Walsh et al. 2008; \S~4.2).

We end with the following note.  The above discussion presupposes that the
optical line emission comes exclusively from stellar ejecta prior to 
their being mixed into the hot phase.  There may be another component.  
Once the ejecta become thermalized, the hot gas, at least in the inner cores
of bulges, should cool and condense on a relatively short timescale.  Recent
{\it Chandra}\ and {\it XMM-Newton}\ observations of ellipticals and bulges
show that the diffuse X-ray-emitting medium typically has temperatures of
$kT \approx 0.3-1$ keV and densities of $n \approx 0.1-0.3$ \cc\ (see Ho 2009
and references therein).  Assuming $kT=0.5$ keV and $n = 0.1$ \cc, the cooling
time is merely $\sim 6 \times 10^7$ yr.  This simple picture of galactic-scale 
cooling flows encounters a number of long-standing difficulties familiar in 
the context of galaxy clusters, and their resolution appeals to similar 
remedies (Mathews \& Brighenti 2003).  A key ingredient to counter the
overcooling problem invokes some form of energy feedback.  It is tempting to
associate the evidence for AGN feedback discussed in this paper (\S~4.2) with
such a mechanism.  The kinematics that ensue from the cooling flow scenario are
likely to be quite complicated, but in general we also expect the cooling 
condensations to be in pressure balance with the surrounding hot medium and 
hence to reflect its kinematics.  Depending on the shape of the potential, the
cooling flow filaments can settle in different planes at different radii
(Tohline et al. 1982).  This may account for the variety of dust structures
often seen in the central regions of early-type galaxies (e.g., Lauer et al. 
2005).

\section{Summary}

We use a new catalog of stellar velocity dispersions to study the relationship 
between the kinematics of the ionized gas and the stars in the central few 
hundred parsecs of a large sample of nearby galaxies.  The gas dispersions are 
based on the width of the \nii\ \lamb6583 line.  Robust measurements of both 
\sigg\ and \sigs\ are available for 345 galaxies spanning a wide range in 
Hubble type and level of nuclear activity.  The principal results of our 
analysis can be summarized as follows.

\begin{enumerate}

{\item The velocity dispersion of the ionized gas strongly correlates with, and 
is roughly comparable to, the velocity dispersion of the stars over the 
velocity range $\sim 30-350$ \kms, across all Hubble types.  This confirms the 
notion that the widths of the narrow emission lines primarily reflect the 
virial velocity of the gravitational potential of the stellar bulge.  In 
detail, $\sigma_g/\sigma_*$ ranges from $\sim 0.6$ to 1.4, with an average 
value of 0.80.}

{\item The ratio $\sigma_g/\sigma_*$ shows no strong dependence on Hubble type,
for galaxies ranging from E to Sbc.  Among galaxies of type Sc and later, 
systems with \sigs\ \lax\ 40 \kms, all spectroscopically classified as \hii\ 
nuclei, \sigg/\sigs\ \gax\ 1. The ionized gas maintains a minimum ``floor'' 
dispersion of \sigg\ $\approx$ 30 \kms, which might signify a lower threshold 
level of interstellar turbulence generated by energy feedback from massive 
stars.  The $\sigma_g$-$\sigma_*$ relation does not depend on the presence of 
a large-scale bar or on the local environment of the galaxy.}

{\item The residuals of the \sigg-\sigs\ relation for AGNs (Seyferts, LINERs, 
and transition objects) correlate strongly with the nuclear optical (H\al) 
luminosity, and perhaps even stronger with the Eddington ratio.  Radio power 
plays a less essential role.  We argue that energy injected by accretion disk 
winds or outflows, not radio jets, provide a source of AGN feedback that 
accelerates the NLR gas to speeds above the virial velocities of the stars.}

{\item We confirm that the line widths of narrow emission lines can be used to 
obtain reasonably reliable estimates of \sigs, but, for maximum accuracy,
preference should be given to low-ionization transitions such as \nii\ and 
care should be taken to correct for the secondary dependence of \sigg\ on AGN 
luminosity or Eddington ratio.}

{\item We show that the mass budget of the ionized gas can be accounted for by 
mass loss from evolved stars.  The optical line emission arises from 
photoionization of the just-shed, mass loss envelopes prior to their being 
thermalized, in conjunction with an additional contribution from filaments 
recondensing from the hot medium.  We argue that the line widths are 
dominated by random motions, reflecting the kinematics of the hot gas that 
pressure confines the line-emitting clouds.}

{\item Lastly, we suggest that the line width-luminosity correlation in AGNs
actually defines the upper envelope of a distribution of points governed by 
the \sigg-\sigs\ relation, the \mbh-\sigs\ relation, and the Eddington limit.}
\end{enumerate}

\acknowledgements
I am grateful to Joel Bregman, Jenny Greene, Janice Lee, and Mark Whittle for 
helpful correspondence.  I thank Bill Mathews, the referee, for constructive 
criticisms that helped to clarify the discussion in \S4.4.  This work was 
supported by the Carnegie Institution of Washington and by NASA grants from 
the Space Telescope Science Institute, which is operated by the Association of 
Universities for Research in Astronomy, Inc., for NASA, under contract 
NAS5-26555.

%



\begin{thebibliography}{}

\bibitem[]{}
Atkinson, J. W., et al. 2005, \mnras, 359, 504

\bibitem[]{}
Baum, S.~A., \& McCarthy, P.~J. 2000, \aj, 119, 2634

\bibitem[]{}
Bertola, F., Buson, L.~M., \& Zeilinger, W.~W. 1992, \apj, 401, L79

\bibitem[]{}
Bertola, F., Cinzano, P., Corsini, E.~M., Rix, H.-W., \& Zeilinger, W.~W. 
1995, \apj, 448, L13

\bibitem[]{}
Bian, W., Gu, Q., Zhao, Y., Chao, L., \& Cui, Q. 2006, \mnras, 372, 876

\bibitem[]{}
Boroson, T. A. 2003, \apj, 585, 647

\bibitem[]{}
------. 2005, \aj, 130, 381

\bibitem[]{}
Botte, V., Ciroi, S., Di Mille, F., Rafanelli, P., \& Romano, A. 2005, \mnras,
356, 789

\bibitem[]{}
Busko, I.~C., \& Steiner, J.~E. 1992, \mnras, 258, 306

\bibitem[]{}
Caldwell, N. 1984, \pasp, 96, 287

\bibitem[]{}
Caldwell, N., Kirshner, R.~P., \& Richstone, D.~O. 1986, \apj, 305, 136

\bibitem[]{}
Caon, N., Macchetto, D., \& Pastoriza, M. 2000, \apjs, 127, 39

\bibitem[]{}
Carollo, C.~M., Stiavelli, M., de Zeeuw, P.~T., \& Mack, J. 1997, \aj,
114, 2366

\bibitem[]{}
Chen, X.-Y., Hao, C.-N., \& Wang, J. 2008, ChJAA, 8, 25

\bibitem[]{}
Cinzano, P., Rix, H.-W., Sarzi, M., Corsini, E.~M., Zeilinger, W.~W., \&
Bertola, F. 1999, \mnras, 307, 433

\bibitem[]{}
Cinzano, P., \& van der Marel, R.~P. 1994, \mnras, 270, 325

\bibitem[]{}
Crenshaw, D.~M., Kraemer, S.~B., \& George, I. M. 2003, \annrev, 41, 117

\bibitem[]{}
Das, V., Crenshaw, D. M., Deo, R. P., \& Kraemer, S. B. 2006, \aj, 132, 620

\bibitem[]{}
Dasyra, K. M., et al. 2008, \apj, 674, L9

\bibitem[]{}
Demoulin-Ulrich, M.-H., Butcher, H.~R., Boksenberg, A. 1984, \apj, 285, 527

\bibitem[]{}
de Vaucouleurs, G., de Vaucouleurs, A., Corwin, H.~G., Jr., Buta, R.~J.,
Paturel, G., \& Fouqu\'e, R. 1991, Third Reference Catalogue of Bright
Galaxies (New York: Springer)

\bibitem[]{}
Dumas, G., Mundell, C. G., Emsellem, E., \& Nagar, N. M. 2007, \mnras, 379, 1249

\bibitem[]{}
Falc\'on-Barroso, J., et al. 2006, \mnras, 369, 529

\bibitem[]{}
Ferrarese, L., \& Merritt, D. 2000, \apj, 539, L9

\bibitem[]{}
Filho, M.~E., Barthel, P.~D., \& Ho, L.~C. 2000, \apjs, 129, 93

\bibitem[]{}
------. 2002, \apjs, 142, 223

\bibitem[]{}
------. 2006, \aa, 451, 71

\bibitem[]{}
Fillmore, J.~A., Boroson, T.~A., \& Dressler, A. 1986, \apj, 302, 208

\bibitem[]{}
Fisher, D. 1997, \aj, 113, 950

\bibitem[]{}
Fukugita, M., Shimasaku, K., \& Ichikawa, T. 1995, \pasp, 107, 945

\bibitem[]{}
Gaskell, C.~M., \& Ferland, G.~J. 1984, \pasp, 96, 393

\bibitem[]{}
Gebhardt, K., \etal 2000, \apj, 539, L13

\bibitem[]{}
Greene, J. E., \& Ho, L. C. 2005a, ApJ, 627, 721

\bibitem[]{}
------. 2005b, ApJ, 630, 122

\bibitem[]{}
------. 2006, ApJ, 641, 117

\bibitem[]{}
------. 2007, ApJ, 667, 131

\bibitem[]{}
Greene, J. E., Ho, L. C., \& Ulvestad, J. S. 2006, ApJ, 636, 56

\bibitem[]{}
Grupe, D., \& Mathur, S. 2004, \apj, 606, L41

\bibitem[]{}
Gu, Q.~S., Melnick, J., Cid Fernandes, R., Kunth, D., Terlevich, E., \& 
Terlevich, R. 2006, \mnras, 366, 480

\bibitem[]{}
Heckman, T.~M., Blitz, L., Wilson, A.~S., Armus, L., \& Miley, G.~K. 1989, 
\apj, 342, 735

\bibitem[]{}
Heckman, T.~M., Illingworth, G.~D., Miley, G.~K., \& van Breugel, W.~J.~M.  
1985, \apj, 299, 41

\bibitem[]{}
Heckman, T.~M., Miley, G.~K., van Breugel, W.~J.~M., \& Butcher, H.~R. 1981, 
\apj, 247, 403

\bibitem[]{}
Ho, L.~C. 1996, in The Physics of LINERs in View of Recent Observations, ed. M. Eracleous et al. (San Francisco: ASP), 103

\bibitem[]{}
------. 2002, \apj, 564, 120

\bibitem[]{}
------. 2007, \apj, 668, 94

\bibitem[]{}
------. 2008, \annrev, 46, 475

\bibitem[]{}
------. 2009, \apj, in press

\bibitem[]{}
Ho, L.~C., Filippenko, A.~V., \& Sargent, W.~L.~W. 1995, \apjs, 98, 477

\bibitem[]{}
------. 1997a, \apjs, 112, 315 

\bibitem[]{}
------. 1997b, \apj, 487, 568

\bibitem[]{}
------. 1997c, \apj, 487, 579

\bibitem[]{}
------. 2003, \apj, 583, 159

\bibitem[]{}
Ho, L.~C., Filippenko, A.~V., Sargent, W.~L.~W., \& Peng, C.~Y. 1997d, \apjs,
112, 391

\bibitem[]{}
Ho, L.~C., Greene, J. E., Filippenko, A.~V., \& Sargent, W.~L.~W. 2009, \apjs, 
in press

\bibitem[]{}
Ho, L.~C., \& Peng, C.~Y. 2001, \apj, 555, 650

\bibitem[]{}
Ho, L.~C., Sarzi, M., Rix, H.-W., Shields, J.~C., Rudnick, G., Filippenko,
A.~V., \& Barth, A.~J. 2002, \pasp, 114, 137

\bibitem[]{}
Ho, L.~C., \& Ulvestad, J.~S. 2001, \apjs, 133, 77

\bibitem[]{}
Isobe, T., Feigelson, E.~D., \& Nelson, P.~I. 1986, \apj, 306, 490

\bibitem[]{}
Jim\'enez-Benito, L., D\'\i az, A.~I., Terlevich, R., \& Terlevich, E. 2000, 
\mnras, 317, 907

\bibitem[]{}
Kent, S.~M. 1988, \aj, 96, 514

\bibitem[]{}
Kormendy, J., \& Kennicutt, R. C.  2004, \annrev, 42, 603

\bibitem[]{}
Kormendy, J., \& Westpfahl, D.~J. 1989, \apj, 338, 752

\bibitem[]{}
Kukula, M.~J., Pedlar, A., Baum, S.~A., O'Dea, C.~P. 1995, \mnras, 276, 1262

\bibitem[]{}
Kuzio de Naray, R., McGaugh, S. S., \& de Blok, W. J. G. 2008, \apj, 676, 920

\bibitem[]{}
Laor, A. 2003, \apj, 590, 86

\bibitem[]{}
Lauer, T.~R., et al. 2005, \aj, 129, 2138

\bibitem[]{}
Malkan, M.~A., Gorjian, V., \& Tam, R. 1998, \apjs, 117, 25

\bibitem[]{}
Martini, P., Regan, M.~W., Mulchaey, J.~S., \& Pogge, R.~W. 2003, \apj, 589, 774

\bibitem[]{}
Mathews, W.~G. 1990, \apj, 354, 468

\bibitem[]{}
Mathews, W.~G., \& Brighenti, F. 2003, \annrev, 41, 191

\bibitem[]{}
Minkowski, R., \& Osterbrock, D.~E. 1959, \apj, 129, 583

\bibitem[]{}
Mundell, C.~G., Wrobel, J. M., Pedlar, A., \& Gallimore, J. F. 2003, \apj,
583, 192

\bibitem[]{}
Nagar, N.~M., Falcke, H., \& Wilson, A.~S. 2005, \aa, 435, 521

\bibitem[]{}
Nelson, C.~H. 2000, \apj, 544, L91

\bibitem[]{}
Nelson, C.~H., \& Whittle, M. 1995, \apjs, 99, 67

\bibitem[]{}
------. 1996, \apj, 465, 96

\bibitem[]{}
Noel-Storr, J., Baum, S. A., \& O'Dea, C. P. 2007, \apj, 663, 71

\bibitem[]{}
Osterbrock, D.~E. 1989, Astrophysics of Gaseous Nebulae and Active
 Galactic Nuclei (Mill Valley: Univ. Science Books)

\bibitem[]{}
Padovani, P., \& Matteucci, F. 1993, \apj, 416, 26

\bibitem[]{}
Parriott, J. R., \& Bregman, J. N. 2008, \apj, 681, 1215

\bibitem[]{}
Phillips, A.~C., Illingworth, G.~D., MacKenty, J.~W., \& Franx, M. 1996, \aj,
111, 1566

\bibitem[]{}  
Phillips, M.~M., Charles, P.~A., \& Baldwin, J.~A. 1983, \apj, 266, 485

\bibitem[]{}
Phillips, M.~M., Jenkins, C.~R., Dopita, M.~A., Sadler, E.~M., \&
Binette, L. 1986, \aj, 91, 1062

\bibitem[]{}
Pignatelli, E., et al.  2001, \mnras, 323, 188

\bibitem[]{}
Pizzella, A., Corsini, E., Sarzi, M., Magorrian, J., Mendez-Abreu, J., 
Coccato, L., \& Bertola, F. 2008a, \mnras, 387, 1099

\bibitem[]{}
Pizzella, A., Tamburro, D., Corsini, E. M., \& Bertola, F. 2008b, \aa, 482, 53

\bibitem[]{}
Pogge, R.~W., Maoz, D., Ho, L.~C., \& Eracleous, M. 2000, \apj, 532, 323

\bibitem[]{}
Proga, D. 2007, in The Central Engine of Active Galactic Nuclei, ed. L. C. Ho 
\& J.-M. Wang (San Francisco: ASP), 267

\bibitem[]{}
Ravindranath, S., Ho, L.~C., Peng, C.~Y., Filippenko, A.~V., \& Sargent, 
W.~L.~W. 2001, \aj, 122, 653

\bibitem[]{}
Rice, M. S., Martini, P., Greene, J. E., Pogge, R. W., Shields, J. C., 
Mulchaey, J. S., \& Regan, M. W. 2006, \apj, 636, 654

\bibitem[]{}
Salviander, S., Shields, G. A., Gebhardt, K., \& Bonning, E. W. 2007, \apj, 
662, 131

\bibitem[]{}
Sanders, R.~H. 1981, \apj, 244, 82

\bibitem[]{}
Sarzi, M., et al. 2006, \mnras, 366, 1151

\bibitem[]{}
Sarzi, M., Rix, H.-W., Shields, J.~C., Rudnick, G., Ho, L.~C., McIntosh,
D.~H., Filippenko, A.~V., \& Sargent, W.~L.~W. 2001, \apj, 550, 65

\bibitem[]{}
Schmitt, J.~H.~M.~M. 1985, \apj, 293, 178

\bibitem[]{}
Shields, G.~A., Gebhardt, K., Salviander, S., Wills, B.~J., Xie, B.,
Brotherton, M.~S., Yuan, J., \& Dietrich, M. 2003, \apj, 583, 124

\bibitem[]{}
Sim\~{o}es Lopes, R. D., Storchi-Bergmann, T., de F\'atima de O. Saraiva, M., 
\& Martini, P. 2007, \apj, 655, 718

\bibitem[]{}
Smith, E.~P., Heckman, T.~M., \& Illingworth, G.~D. 1990, \apj, 356, 399

\bibitem[]{}
Terashima, Y., \& Wilson, A.~S. 2003, \apj, 583, 145

\bibitem[]{}
Terlevich, E., D\'\i az, A. I., \& Terlevich, R. 1990, \mnras, 242, 271

\bibitem[]{}
Tohline, J.~E., Simonson, G.~F., \& Caldwell, N. 1982, \apj, 252, 92

\bibitem[]{}
Tremaine, S., et al. 2002, \apj, 574, 740

\bibitem[]{}
Ulvestad, J.~S., \& Ho, L.~C. 2001, \apj, 558, 561

\bibitem[]{}
Vega~Beltr\'an, J.~C., Pizzella, A., Corsini, E.~M., Funes, J.~G.,
Zeilinger, W.~W., Beckman, J.~E., \& Bertola, F. 2001, \aa, 374, 394

\bibitem[]{}
Veilleux, S. 1991, \apjs, 75, 383

\bibitem[]{}
Verdoes Kleijn, G.~A., van der Marel, R.~P., \& Noel-Storr, J. 2006, \aj, 
131, 1961

\bibitem[]{}
V\'eron, M.-P. 1981, \aa, 100, 12

\bibitem[]{}
Vrtilek, J.~M., \& Carleton, N.~P. 1985, \apj, 294, 106

\bibitem[]{}
Walsh, J. L., Barth, A. J., Ho, L. C., Filippenko, A. V., Rix, H.-W., Shields, 
J. C., Sarzi, M., \& Sargent, W. L. W. 2008, \aj, 136, 1677

\bibitem[]{}
Wang, T.-G., \& Lu, Y.-J. 2001, \aa, 377, 52

\bibitem[]{}
Whittle, M. 1985, \mnras, 213, 1

\bibitem[]{}
------. 1992a, \apjs, 79, 49

\bibitem[]{}
------. 1992b, \apj, 387, 109

\bibitem[]{}
------. 1992c, \apj, 387, 121

\bibitem[]{}
------. 1993, in The Nearest Active Galaxies, ed. J. Beckman,
L. Colina, \& H. Netzer (Madrid: CSIC Press), 63

\bibitem[]{}
Whittle, M., \& Wilson, A.~S. 2004, \aj, 127, 606

\bibitem[]{}
Wilson, A.~S., \& Heckman, T.~M. 1985, in Astrophysics of Active Galaxies
and Quasi-Stellar Objects,  ed. J.~S. Miller (Mill Valley, CA: Univ. Science
Books), 39

\bibitem[]{}
Zhou, H.-Y., Wang, T.-G., Yuan, W., Lu, H., Dong, X., Wang, J., \& Lu, 
Y. 2006, \apjs, 166, 128

\end{thebibliography}
\end{document}